\documentclass[sigconf,10pt,nonacm]{acmart}
\usepackage{xspace}

\newcommand{\eg}{\emph{e.g.,}\xspace}

\title{The Frontier LLM Trap in Network Automation}

\newcommand{\remove}[1]{}

\newcommand{\myitem}[1]{\noindent\textbf{#1}}
\newcommand{\sys}{NetMentor\xspace}

\author{Minhao Jin}
\affiliation{%
  \institution{Princeton University}%
  \city{Princeton, NJ}
  \country{USA}%
}
\author{Sean Wang}
\affiliation{%
  \institution{Princeton University}%
  \city{Princeton, NJ}
  \country{USA}%
}
\author{Aarti Gupta}
\affiliation{%
  \institution{Princeton University}%
  \city{Princeton, NJ}
  \country{USA}%
}
\author{Maria Apostolaki}
\affiliation{%
  \institution{Princeton University}%
  \city{Princeton, NJ}
  \country{USA}%
}
\usepackage{multirow}
\usepackage{mathtools}
\usepackage{xspace}
\usepackage{xcolor}
\usepackage{multirow}
\usepackage{graphicx}
\usepackage{booktabs}
\usepackage{tabularx}
\usepackage{array}
\usepackage{caption}
\usepackage{subcaption}
\usepackage{amsmath}
\usepackage{bm}
\usepackage{pifont}% http://ctan.org/pkg/pifont
\usepackage{url}
\usepackage{xurl}
\usepackage{makecell}
\usepackage{adjustbox}
\usepackage{enumitem}

\usepackage{caption}

\usepackage{algorithm,algpseudocode}
\algnewcommand{\To}{\textbf{To }}
\algnewcommand\Input{\item[\textbf{Input:}]}%
\algnewcommand\Output{\item[\textbf{Output:}]}%
\algrenewcommand{\algorithmiccomment}[1]{%
  \hfill\(\triangleright\)\textit{\tiny #1}%
}

\usepackage{booktabs} % Add this to your preamble
\usepackage{multirow}

\usepackage{mathtools}

\usepackage{hyperref}

\newcommand{\minhao}[1]{{\footnotesize\color{blue}[MJ: #1]}}

\newcommand{\ag}[1]{{\footnotesize\color{green}[AG: #1]}}

\newcommand{\todo}[1]
{{\footnotesize\color{red}[ToDo: #1]}}

\newcounter{packednmbr}

\newcommand{\mypara}[1]{\smallskip \noindent{\bf {#1}.}~}

\newcommand{\OMIT}[1]{{}}

\newcommand{\gramcfg}{\langle \text{cfg} \rangle}
\newcommand{\gramparam}{\langle \text{param} \rangle}
\newcommand{\gramtype}{\langle \text{type} \rangle}
\newcommand{\gramstmt}{\langle \text{stmt} \rangle}
\newcommand{\gramline}{\langle \text{line} \rangle}
\newcommand{\gramctx}{\langle \text{ctx} \rangle}
\newcommand{\gramctxline}{\langle \text{ctx\_line} \rangle}
\newcommand{\grampred}{\langle \text{P} \rangle}
\newcommand{\gramcond}{\langle \text{C} \rangle}
\newcommand{\gramaction}{\langle \text{A} \rangle}
\newcommand{\actinject}{\textbf{inject}}
\newcommand{\actreplace}{\textbf{replace}}
\newcommand{\actdrop}{\textbf{drop}}
\newcommand{\match}{\texttt{match}}

\newcommand{\qwenmodel}{Qwen2.5-Coder-32B-Instruct }

\usepackage{ifthen}

\usepackage{listings}
\newcounter{insightlabel}
\newcounter{insightnmbr}
\renewcommand{\theinsightlabel}{\textbf{\theinsightnmbr}}

\newcounter{challengelabel}
\newcounter{challengenmbr}
\renewcommand{\thechallengelabel}{\textbf{\thechallengenmbr}}

\newcounter{researchquestionlabel}
\newcounter{researchquestionnmbr}
\renewcommand{\theresearchquestionlabel}{\textbf{\theresearchquestionnmbr}}

\usepackage{mdframed}

\newcounter{findinglabel}
\newcounter{findingnmbr}
\renewcommand{\thefindinglabel}{\textbf{\thefindingnmbr}}

\newcounter{limitationlabel}
\newcounter{limitationnmbr}
\renewcommand{\thelimitationlabel}{\textbf{\thelimitationnmbr}}

\newcommand{\commentout}[1]{}

\newcommand{\rom}[1]{\uppercase\expandafter{\romannumeral #1\relax}}

\usepackage[subtle]{savetrees}
\begin{abstract}

Large LLMs are powerful tools for network automation, but they are expensive, slow to serve, hard to audit, poorly tailored to individual networks, and create long-term dependencies on a small number of AI providers. 
Existing alternatives fall short: small open-source models are cheaper but unreliable, while deterministic scripts and verification are controllable but hard to build and maintain.

We propose a middle ground. An offline loop composed of 
fuzzing and validation discovers the recurring mistakes small models make, then resulting networking knowledge is expressed as explicit logic rules. In production, these rules guide a small model on tasks such as configuration translation, yielding automation that is cheaper, lower-latency, auditable, and easier to adapt to a specific network.

More critically, network knowledge and operational experience stay and are accumulated where they belong, the network itself, not a rented service.

\end{abstract}

\begin{document}

\maketitle

\section{Introduction}

Networking tasks are both difficult and diverse, spanning configuration, troubleshooting, verification, performance tuning, and policy enforcement. Traditional automation tools struggle to keep up: they often require substantial manual engineering, must be rewritten for each new task or network, and are difficult for operators to adopt, maintain, and extend.

Frontier LLMs offer a compelling alternative~\cite{openai,claude,germini,xai}, promising enormous speed-ups for many of these tasks, but have significant limitations in practice. 
First, they are expensive, often impossible to serve, making them especially hard to justify for smaller networks. 
Second, they are often slow. We find that GPT-5.5 \cite{openai2026gpt55} can require up to four attempts to produce a correct configuration translation, with each attempt taking as long as 2.5 minutes.
% \todo{GPT-5 can take up to x to answer to config translation}.
%Third they often make mistakes on-shot, requiring additional human led refinement. We find that \todo{GPT-5 we correct only in 3/113} experiments.
Finally, they also create a fragile operational dependency on a small number of AI providers: providers may change prices, restrict usage, degrade access, or disappear altogether. As a result, industry is increasingly looking beyond frontier-model-only solutions toward cheaper, more controllable, and more deployable forms of network automation~\cite{ms_news}. %\maria{maybe add satya's letter?}.

This paper asks: How can we build a network automation agent that helps operators with tasks such as configuration translation, troubleshooting, verification, and policy enforcement, while being effective, fast, controllable, and tailored to a specific network? More importantly, how can such an agent improve over time without creating a long-term dependency on frontier LLM providers?

A natural starting point is to use a small, inexpensive, locally deployable LLM~\cite{hui2024qwen2,llama,gemma} as the interface for network automation, and then teach it the networking expertise it is missing. Such models can be hosted using the operator’s own infrastructure, making them an attractive option: they are fast, cost-efficient, easy to deploy close to the operator or the network, and give the operator full control over the model and its serving environment.
% \todo{update this based on our definition of small model}. 
% But it immediately raises a chicken-and-egg problem. If we already knew exactly what the model needed to learn, we could encode that knowledge into a deterministic algorithm.
However, determining what exactly to teach the model is no easy task.
In practice, the relevant expertise is often implicit, scattered across vendor behavior, operator experience, incomplete documentation, and the quirks of a specific network.

Training such a model directly is also difficult. Networking lacks large, clean datasets of specific queries paired with correct answers. Even if we could generate queries and candidate answers, correctness is hard to verify, and finding the right answer is non-trivial.  More importantly, it is unclear how to reliably teach a model to avoid similar, or even identical, mistakes. Fine-tuning can be opaque and brittle~\cite{li2024revisiting,luo2025empirical}; prompt engineering creates a context problem and is not reliable~\cite{wu2025lifbench,qi2025agentif,jaroslawicz2025many}, while constrained decoding~\cite{he2025justintime,he2025making,dong2025xgrammar,ugare2024syncode,zhang2023don}, which restricts the model's token choices to what a grammar permits, can overconstrain the model (even preventing it from generating the correct answer) and make the system harder to maintain~\cite{banerjee2025crane,reddy2026draft}. Attempts in the networking domain are useful~\cite{he2025making,he2025justintime} but require the rule grammar to align with tokenization.

We argue for a different approach: instead of trying to teach everything at once, teach small models by systematically discovering and correcting their own recurring mistakes. The key idea is to build an offline teaching loop that combines multiple complementary tools: fuzzers to trigger failures in the small model, frontier LLMs to propose candidate answers and explanations, and a validation engine composed of emulators to test behavior in realistic network settings and/or verifiers to check semantic correctness when possible. Each component is insufficient on its own. Frontier LLMs may hallucinate or propose answers that are incorrect, unnecessary, or inappropriate for a particular network; verifiers can be slow or incomplete; emulators are limited by their fidelity; and fuzzers only reveal the failures they can generate. But together, they can continually expose mistakes, infer missing networking principles, and validate candidate corrections while building on an internally owned model.

This approach is not traditional model distillation~\cite{hsieh2023distilling, mukherjee2023orca,jiang2023lion,zhou2024universalner,wang2023self}. We do not blindly force the small model to imitate a frontier LLM’s outputs. In fact, we find that this strategy would be suboptimal for networking. First, many networking tasks admit multiple correct answers: for example, several configurations may implement the same intended behavior, and over-constraining the model to match one answer can unnecessarily reduce flexibility. Second, correctness is often contextual or network-specific: a solution that is valid in general may be irrelevant or wrong in a network where a particular protocol, feature, or policy is disabled. Third, frontier models themselves are not reliable ground truth; they can hallucinate, omit important constraints, or produce incorrect configurations. We have experienced all of these limitations firsthand.
GPT-5.5 has < 5\% one-shot accuracy for configuration translation, and when it finds a correct answer, it often differs from what a human would write, or has consequences that are not yet visible.
Thus, we use frontier LLMs not as oracles to imitate, but as hypothesis generators whose outputs must be tested, validated, and converted into explicit rules before they guide the small model. 
% \maria{I would remove the example bellow and fill in above? } 
% Consider the task of translating a Cisco OSPF configuration where an interface's OSPF priority is omitted, implicitly relying on Cisco's default value of 1.
% A frontier LLM, or even an experienced engineer, is likely to preserve this omission when translating the configuration to Junos, where the default value is 128. This discrepancy can cause behavioral inconsistencies between the two configurations.
% However, the frontier model can easily correct this misconfiguration by specifying the priority explicitly once it receives related contextual feedback.

%configuration that looks wrong in the beginning might make sense once completed \maria{need to say this better, we need some examples later on to justify this}; there might be multiple right answers; some correct in general but unnecessary for this network, or only one of several valid behaviors. 
%Instead, we use the frontier LLM as a hypothesis generator inside a loop with fuzzing, verification, and emulation.
We then use traditional tools the networking community has developed and relied on for years---verifiers, emulators, and tests---to separate useful hypotheses from plausible but incorrect ones. The validated knowledge is captured as explicit logic rules and applied at inference time, rather than buried inside model weights. In this way, frontier LLMs act as capable but fallible temporary teachers, not permanent dependencies or sources of authority. The resulting loop incrementally builds a network-specific layer of auditable, editable rules that improves a small model over time while keeping the system inexpensive, controllable, and, more importantly, internally owned.

This paper explores the agenda above through the lens of a particular networking task: configuration translation. We ask whether small models can be made practical not by scaling them up, but by surrounding them with the right symbolic knowledge. More broadly, we argue that the next generation of network automation should not be defined by how often we call frontier models, but by how effectively we can extract, validate, and reuse their expertise.
\begin{figure*}[h]
\centering
\includegraphics[width=\linewidth]{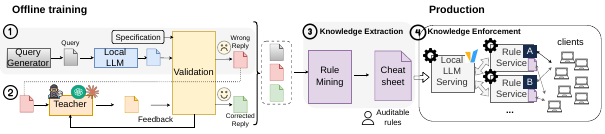}
\caption{%\maria{Change Data Generator to maybe Query? add some notation to file icon e.g. query, reply wrong reply, corrected reply? Robustification or off-line training? it is not clear that knowledge extraction is on robustification } \minhao{Interchange the red and green with feedback to teacher}}
% An overview of the \sys architecture. Offline training consists of (1) data generation and initial generation with a small, local LLM, followed by (2) validation and correction by a teacher model in a feedback loop. The original query and incorrect and corrected responses are used to (3) mine a set of logical rules, which are then (4) enforced for a local LLM during generation.
Offline, \textcircled{1}the query generator produces test queries, the local LLM replies, and validation checks each against the specification; \textcircled{2}once failed, the teacher and validation engine propose a fix; \textcircled{3}the resulting ⟨query, wrong reply, corrected reply⟩ triplet feeds Rule Mining, which extracts auditable rules. In production, \textcircled{4}a per-task rule service applies these rules between the LLM serving engine and end users, with all tasks sharing one serving engine.
}
% \vspace{-1em}
\label{fig:overview}
\end{figure*}

\section{Network Automation Requirements \& Challenges}

To better understand the problem, we put ourselves in the shoes of a network engineer tasked with translating configurations from one vendor or configuration language to another. We proceed by trying two approaches.  
% \maria{can we turn this into what we did vs what a hypothetical engineer would see??}
%The company encourages engineers to use LLMs to improve productivity, and the engineer has two natural options.

\myitem{Approach 1: Frontier LLM with Per-Query Validation.}
We first ask the frontier LLM to translate each configuration, then manually inspect the result and check it using existing tools. At first glance, 
%this will work surprisingly well: her productivity spikes, and routine translations become much faster. 
the generated translations appear plausible, but closer inspection reveals
%the cracks soon appear. The model often adds 
unnecessary configuration statements that preserve behavior but may complicate future network management.
%confuse future operators: why is this line here, and is it safe to remove?  
For example, when translating a Cisco OSPF routing configuration to Junos we found that GPT-5.5 often added static routes in the translated version even though they did not exist in the original. 

Using an emulator, we evaluate the correctness of translations, examining router behaviors such as OSPF attributes, LSA metrics, inter-area routes, and static routes.
We observe that 
%the frontier model frequently produces incorrect translations with similar mistakes. 
the frontier model alone is unreliable: many responses are incorrect, exhibiting similar mistakes, others are correct but contain unnecessary statements for some networks, and many tasks admit multiple valid translations. 
% Our experiments show that the GPT-5.5 achieves < 5\% one-shot accuracy when translating Cisco configurations to a behaviorally equivalent Junos configuration.
% \sw{We already mention this in the intro. Can we cut this from one of these two spots?}
In our experiments, 18.6\% of the generated Junos configs fail to even establish connectivity with emulated test routers.
Hence, we re-prompt GPT-5.5 with the failure context and ask it to propose a minimal correction.
While full validation is achieved within 1.79 repair iterations on average, 
%many errors are easy to fix with few iterations of refinement, 
some corrections take an unpredictable amount of time, often resulting in an ad hoc loop of inspecting the output, modifying the prompt, rerunning the model, and checking again. 

Furthermore, as network size increases, token costs and serving latency increase, reducing productivity gains.
% We also observe that the model often makes similar mistakes.
Worse, the workflow now depends on continued access to a frontier model provider whose pricing, rate limits, policies, or availability may change without the network operator’s control.
 %For example, the ban of Claude Fable 5, the retirement of multiple OpenAI GPT models and the frequent API pricing changes may force operators to transition to other LLMs, which is likely to cause performance differences~\cite{}. 

 % \maria{we need to add some examples of running this here to show that was indeed the case and it is not a hypothetical scenario we made up}

\myitem{Approach 2: Small Model with Logged Experience.}
A natural response is to preserve the frontier model’s outputs and pass them on to a small model. 
We log useful prompts, answers, validation failures, corrections, and examples from previous translations. In effect, we build an internal knowledge base of networking principles and “model operating procedures” for configuration translation. However, this approach also breaks down. The accumulated context quickly becomes too large to include in every prompt, and deciding which pieces of past knowledge are relevant to a new configuration is itself difficult. Even when the right hints are included, a small model may fail to follow them reliably.
Indeed, the LLM community has reported significant degradation of instruction following for LLMs as the context grows~\cite{wu2025lifbench,qi2025agentif,jaroslawicz2025many}. Our experiment also shows that conveying knowledge through prompts lacks guarantees and is 6\% less accurate than strict enforcement even with just 15 short natural-language rules.
% \maria{did we also observe that?} 
% \maria{move what is bellow to the previous myitem and integrate say here that logging everything still does not save us from mistakes}
% More fundamentally, copying the frontier model’s answers is not enough: some answers are wrong, some are correct but unnecessary for this network, and many tasks admit multiple valid translations. 
% Our experiments show that the GPT-5.5 achieves < 5\% one-shot accuracy when translating Cisco configurations to a behaviorally equivalent Junos configuration. Moreover, 18.6\% of the generated Junos configs fail to establish connectivity with the FRR test routers.

%What is needed is not a growing log of examples, but a way to extract the underlying networking principles, validate them, and make them usable by a smaller model.
% \minhao{Add "how many loops to make it succeed in full". 
% X / 100 cannot establish the connection.
% Y/100 create behavior difference. 
% The worst latency we observe from the configs translation.}
% \maria{any examples here?}

\myitem{A better approach:}
Our experiment shows that we need a way to capture the productivity benefits of LLMs without turning frontier models into a permanent production dependency. 
A practical network automation agent should satisfy five requirements.
First, it should be \textbf{effective}: it must produce outputs that are mostly correct and useful for real networking tasks, not merely plausible. Second, it should be \textbf{network-specific}: its behavior should adapt to the protocols, policies, vendors, and operational conventions of the target network. Third, it should be \textbf{low-latency} and \textbf{inexpensive}: operators should be able to use it interactively, and smaller networks should be able to deploy it without paying frontier-model serving costs on every query. Fourth, it should be \textbf{controllable} and \textbf{auditable}: when the agent makes a recommendation, operators should be able to inspect the reasoning, identify the rules being applied, and override or update them when needed. Finally, it should be \textbf{maintainable}: as the network changes, improving the agent should not require rewriting brittle scripts, manually curating ever-growing prompts, or repeatedly fine-tuning opaque model weights.
In short, we want an agent that feels like an LLM to the operator, behaves like a network-specific tool, and can improve over time without locking the organization into a frontier model provider.

\section{Knowledge Extraction 
\& Enforcement Architecture}

Our experience with translating configurations motivated a key change: we need to separate the validation from the inference path. With this idea in mind, we propose  \sys shown in \figureautorefname~\ref{fig:overview}.
\sys splits the expensive per-query validation into: \emph{(i)} an offline loop that identifies concise, reusable, and auditable task-specific rules; and \emph{(ii)} an online stream-like service that enforces them. %\todo{improve description of ii} 
This section explains why each stage is hard and what insights make it tractable and general for multiple networking tasks.

%In this section, we explain why satisfying the above requirements is hard and how our proposed framework, \sys shown in Fig.~\ref{} aims to address them. 

\myitem{Teaching by Failure, Not by Specification.}
Teaching a model network automation from scratch is difficult for the same reason that fully deterministic solutions are difficult: the required knowledge is implicit, incomplete, and potentially unbounded. There is no comprehensive specification of how every network task should be solved, nor a large dataset of inputs paired with correct outputs to train on or analyze. Even human network engineers do not begin with complete knowledge; they start with general principles and learn the details of a particular network through experience.
We adopt the same philosophy for small models. Rather than trying to teach the model everything, we start with an already usable small model and focus on correcting its observed mistakes. 
While intuitive, this approach is hindered by three challenges: \emph{(i)} how do we expose mistakes, given the space of possible user queries? \emph{(ii)} how do we know when a small model is actually wrong? and if it is, \emph{(iii)} 
%even if we know the current answer is wrong, 
how do we fix it without ground truth? 
To address these challenges, %we rely on three key insights:
we constrain the search space to a single task, \eg configuration translation, rather than trying to build a universal networking model. Within a task, the space of possible queries is structured and can be navigated systematically using a fuzzer. While fuzzing can expose suspicious or inconsistent model behavior, it does not by itself tell us whether the output is actually wrong, why it is wrong, or what the correct behavior should be. Our insight is that networking already has a rich ecosystem of reliable tools that can provide this missing signal. Verifiers can check syntactic and some semantic properties~\cite{batfish, campion, minesweeper}, while emulators can test behavior under realistic conditions~\cite{containerLab}. %Used together, these tools can help distinguish real mistakes from harmless variation. 
When the small model’s output is found to be wrong, we rely on a frontier LLM to propose candidate corrections using feedback from the validation step. Observe here that we do not treat the frontier model as an oracle; its suggestions are hypotheses that are passed through the same validation loop
%, using verifiers, emulators, and tests 
to decide whether they capture real networking knowledge. While frontier LLMs are not oracles, we find they are great at repairing their own mistakes with guidance. This behavior might mean that they have the knowledge they need (\eg have been trained on the appropriate documents) but are unlikely to invoke it directly at inference without guidance.  

%to generate diverse task instances, stress the small model, and reveal recurring mistakes. These failures then become the raw material for teaching: instead of asking what the model should know in the abstract, we discover what it needs to know by observing where it fails.

%\myitem{Using Frontier Models Without Outsourcing Ground Truth.}

%\maria{right now we miss some insight into how pairs of wrong-right answers are turned into rules, there is a lot in 4.1 what can we lift?} \ag{Is this level sufficient for sketching the two insights? We can change "insights" --> "design decisions" as needed} \\

\myitem{Capturing Knowledge through Logical Rules.}
By repeatedly asking a frontier LLM to correct outputs based on validator feedback, we obtain a corpus of errors and their corresponding corrections, which we call the \emph{rule mining dataset}. 
We use this corpus to extract logical rules that capture the networking knowledge needed for the task at hand. Specifically, the rules describe \emph{possible repair actions}, %such as adding, dropping, and replacing configuration lines, 
as well as the \emph{conditions under which these actions should be applied}. 
% \maria{maybe add here something like: While intuitive, naively extracting rules from this dataset will fail miserably. Extracted rules would never be useful because they will be too specific.} 
A central challenge is that 
%there may be too many corrections in the rule mining dataset to enforce each one individually, and even if one could, the set of observed errors and repairs 
rules extracted naively from this dataset will fail to generalize to new erroneous outputs, such as network configurations with similar statements, but with different values configured for the parameters. 
% \maria{I think this might read as if the number of rules is the problem, could we emphasize more generality} 
%Therefore, we extract networking principles as logical rules that can be used to impose constraints on LLM generation. Specifically, In other words, Thus, 
To keep the set of rules manageable and generalizable, 
%we require that rules are generalizable and apply in many instances. Our 
our insight is to use a \emph{symbolic representation} for each learned rule, such that it can apply in many instances with possibly different values for the symbolic parameters. 
Rules also need to be auditable and efficiently verifiable, i.e., they should be easy to understand and should provide wide coverage over common cases. %In addition to representing rules symbolically, 
Our second insight is to propose three \emph{domain-specific rule templates} based on observed LLM mistakes. The symbolic representations and rule templates are described in \S\ref{section:rule_design}. We learn the rules manually in our current prototype, and plan to automate this using LLMs (guided by synthesis) in future work. 

\myitem{Enforcing Knowledge with a Streaming Service.}
Enforcement must be lightweight, or local deployment forfeits its cost advantage; reliable, in that no returned response violates the active rule set; and minimally invasive, treating the model and serving stack as black boxes.
Prompting fails on reliability: instructions are advice, and following degrades as context grows ~\cite{wu2025lifbench,qi2025agentif,jaroslawicz2025many}. Constrained decoding~\cite{he2025justintime,he2025making,dong2025xgrammar}, which masks each next-token distribution to what a grammar permits, is reliable by construction but puts rule logic inside the decoding loop, and cannot express rules that are not token-local, i.e., rules whose validity depends on tokens generated much later.
% The latter is particularly damaging because many networking rules are not token-local. 
In Cisco-to-Junos translation, for example, the small model (\eg Qwen2.5-Coder-32B-Instruct~\cite{hui2024qwen2}) may generate an OSPF \texttt{export} statement before generating the corresponding \texttt{policy-options} block. A rule such as “\texttt{drop 'export' if no 'policy-options' exists}” cannot be decided when the token `export` first appears, because the required block may be generated later. 
Thus, token-level decoding would either reject valid intermediate outputs too early or require complex lookahead and global state.
We therefore enforce rules with a streaming chunk-repair proxy that operates over completed configuration blocks rather than individual tokens, sidestepping token-locality entirely (see~\S\ref{subsec:rule_enforce}).

% Therefore, we propose a streaming-based chunk repair mechanism. Instead of injecting rule logic into the LLM or serving stack, our approach runs a separate Rule Service that stores and enforces task-specific rules with minimal invasion. The Rule Service is interposed between the user and the LLM service: it forwards the user request to the LLM, listens to the streamed response, and incrementally validates the generated text.

% Rule enforcement does not wait only until the full answer generated. During streaming, the Rule Service buffers tokens into semantically meaningful chunks, such as completed Junos configuration blocks. The completion of such a chunk triggers the checker: the proxy parses or matches the chunk against the active rule set.

% When there is violation, the enforcement stops the streaming and directly repair the generated text, not by changing model weights, logits, or token probabilities. It then resends the original prompt together with this fixed prefix to the LLM service so generation can continue from the corrected state. Thus, the proxy enforce the rules through output-level repair and controlled regeneration, while the LLM and serving stack remain black boxes.

% This design avoids token-level search by applying compact rule-guided repairs at chunk boundaries. Since regenerated requests reuse the repaired prefix, prefix/KV-cache reuse can further reduce restart overhead when supported. The system therefore returns outputs that satisfy all checked rules while preserving lightweight, black-box serving.

\OMIT{
Instead, we propose a streaming-based chunk repair mechanism. \sys runs a separate Rule Service as a proxy between the user and the LLM service, leaving the LLM and serving stack unchanged. The proxy forwards the request to the LLM, listens to the streamed response, and buffers tokens into semantically meaningful chunks, such as completed Junos configuration blocks. Each completed chunk triggers the checker, which parses or matches the chunk against the active task-specific rules.
When a violation is detected, the proxy stops the current stream and repairs the generated text directly. \sys does not change model weights, logits, or token probabilities; neither does it waste resources by waiting for a full response and requesting from scratch. 
Instead, it treats the repaired text as a fixed prefix and resends the original prompt with this prefix to the LLM service, allowing generation to continue from the corrected state. This avoids token-level search while enabling prefix/KV-cache reuse when supported. 
}
%Before returning the response, the system ensures that the final output satisfies all checked rules while preserving lightweight, black-box serving.

\begin{figure}[t]
\centering
\includegraphics[width=.9\linewidth]{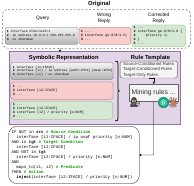}
\vspace{-1em}
\caption{%\minhao{Drop left side of the subfigure. Add a label for the note and replace it with config}
An example of rule mining for configuration translation. Rule Mining converts the ⟨query, wrong reply, corrected reply⟩ triplet to symbolic form, matches it to a rule template, and captures it as a logical repair rule.}
\label{fig:config_translate}
\vspace{-1em}
\end{figure}

\section{Initial Design}

\subsection{Rules for Configuration Translation}\label{section:rule_design}

We start by describing the symbolic representations of configurations and rules, and then present three useful templates along with an illustration on the example in \figureautorefname~\ref{fig:config_translate}.

\begin{figure}[t]
    % \begin{subfigure}{0.8\linewidth}
    \begin{small}
        \begin{align*}
        \textbf{Configurations} & \\
            \gramcfg &\Coloneqq \gramctxline^+ \\
            \gramctxline &\Coloneqq \gramctx\gramline \\
            \gramline &\Coloneqq  (\gramstmt \,|\, \gramparam)^+ \\
            \gramctx &\Coloneqq \text{emp} \,|\, (\gramline/)^+ \\
            \gramstmt &\Coloneqq s \in S \\
            \gramparam &\Coloneqq v : \gramtype \\ 
            \gramtype &\Coloneqq \text{NUM} \,|\, \text{IPV4} \,|\, \text{PFX4} \,|\, \text{IFACE} \,|\, \text{AREA\_ID} \\\\
            \textbf{Rules} & \\
            \text{Conditions} \; \gramcond &\Coloneqq \gramctxline \,|\, \neg \gramctxline \\
            \text{Predicates} \; \grampred &\Coloneqq \textbf{true} \,|\, \textbf{false} \,|\, v_1 = v_2 \,|\, R \\ 
                &\,|\, \neg\grampred \,|\, \grampred \land \grampred \,|\, \grampred \lor \grampred \\
           \text{Actions} \; \gramaction &\Coloneqq \actinject(\gramctxline) \,|\, \actdrop(\gramctxline) \\
                &\,|\, \actreplace(\gramctxline, \gramctxline)
        \end{align*}
    \end{small}
    % \end{subfigure}
    \vspace{-1em}
    \caption{Grammar for Symbolic Representation of Configurations and Rules} %\minhao{add "template"}}
    \vspace{-1em}
    \label{fig:grammar}
\end{figure}

\OMIT{
During generation, configuration correctness is bolstered by enforcing logical constraints on partial outputs given by the model. In order to effectively capture the frontier model's configuration refinement capabilities during the teaching phase, these rules need to describe \emph{possible repair actions}, such as adding, dropping, and replacing configuration lines, as well as the \emph{conditions under which these actions should be applied}. Rules also need to be auditable and efficiently verifiable, so rather than allowing arbitrary logical formulas as rules, we provide templates tailored to common mistakes encountered during configuration translation. \ag{relate this to task-based specifications later} 
%% why symbolic?
Furthermore, these rules are symbolic, meaning that configuration parameters are represented by typed symbolic variables that can take different values, so that the same rules can be applied in multiple instances.
}

%% why context?
\OMIT{Previous works for learning relationships between configuration lines treat configurations as unstructured text and extract abstract patterns, avoiding the need to recognize vendor-specific syntax and enabling identification of similar lines~\cite{}\todo{mention concord here?}. We adopt a similar approach to identify lines with similar roles. Specifically, we associate a line with a context, which identifies its position in the hierarchy of the configuration.
\todo{explain why symbolic lines + context are good for repair}}

\OMIT{
\begin{figure}[h]
    \centering
    \includegraphics[width=\linewidth]{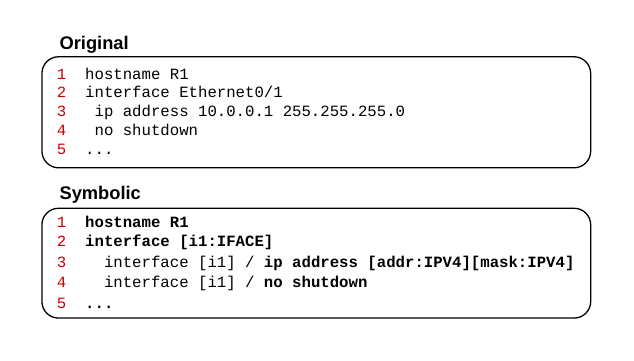}
    \caption{Caption}
    \label{fig:grammar_example}
\end{figure}
}

%\ag{give an example first, and then illustrate the grammar on it}

\myitem{Symbolic Grammar for Configurations.} 
%\maria{is this for the fuzzer or for understanding the output of the LLM?}
 To symbolically represent configurations, we adopt a similar approach to previous work~\cite{concord, configv}, which uses abstract patterns to identify similar lines. The full grammar for symbolic configurations and rules is shown in \figureautorefname~\ref{fig:grammar}. Configurations, $\gramcfg$, are sequences of lines with contexts (context-lines), which consist of statements and typed parameters. %Configuration lines provide a syntactic pattern for comparing lines, while 
 Contexts describe the position of a line in the configuration's hierarchy. 
 
 To illustrate, consider the query in \figureautorefname~\ref{fig:config_translate}, which contains part of a Cisco configuration. In the symbolic representation, the interface name, IP address, and mask are extracted as parameters $i_1, \textit{ addr}$, and $\textit{mask}$, respectively. Also, lines~2 and 3 %Certain lines also 
 are associated with the context "\texttt{interface [i1]}", indicating that they configure the interface $i_1$ in the hierarchy.

\OMIT{To illustrate our symbolic grammar for mining rules, consider the example shown in \figureautorefname~\ref{fig:grammar_example}, which shows part of a Cisco IOS configuration and its corresponding representation. This snippet configures a router, "\texttt{R1}", defines an interface, "\texttt{Ethernet0/1}", and sets the address and "\texttt{no shutdown}" option. In the symbolic grammar, the interface name, IP address, and mask are converted to parameters $i_1, addr, mask$ with types IFACE and IPV4, respectively. Moreover, the interface definition serves as the context for the two following lines, indicating the hierarchical relationship between them. 

A symbolic configuration line consists of configuration statements and the typed parameters of these statements, providing a syntactic pattern for comparing lines. On the other hand, a context consists of a possibly empty sequence of lines denoting the sequence of parents of a given line in the configuration's hierarchy. Together, these components capture both the syntax and structure of a given network configuration.}

%% what is the structure of a rule?
\myitem{Rules for Repair.}
A rule consists of three parts: (1) match conditions on context-lines, which capture when a rule is applied, (2) parameter predicates, which capture the relationships between symbolic variables within or across lines and contexts, and (3) a repair action, which describes how to fix an error in the target configuration.
%addition, deletion, or replacement of a line in the target configuration to fix an error.

%, and serve as the basis for defining the conditions under which repair rules should be triggered. 

%% Role of match conditions
\myitem{Match Conditions.}
We say that two symbolic lines $L_1$ and $L_2$ match, denoted $\match(L_1, L_2)$, if they describe the same pattern of statements and parameters.
% $L_1$ and $L_2$ contain the same number of statements and parameters, where statements in corresponding positions are equivalent (i.e., the same text), and parameters in corresponding positions have the same type. 
For instance, the line "$\texttt{interface } [i_1:\text{IFACE}]$" from \figureautorefname~\ref{fig:config_translate} can only match other lines of the form "$\texttt{interface } [x:\text{IFACE}]$," where $x$ is any variable name.
% and the following lines:
\OMIT{
\begin{align*}
    &l_1: \texttt{interface } [i_2:\text{IFACE}] \\
    &l_2: \texttt{interface } [n:\text{NUM}] \\
\end{align*}

$L$ matches $l_1$, but not $l_2$. 
}
To find matches for similar lines in specific contexts, we extend matching to context-lines, and we say that $\match((C_1, L_1), (C_2, L_2))$ holds if $\match(L_1, L_2)$ and corresponding lines in $C_1$ and $C_2$ match.
\OMIT{
The match conditions of rules can be applied to configurations 
%represented by a sequence of lines with contexts, 
in two ways.
}
For a given context-line $(C, L) \OMIT{\in \gramctxline}$, a match condition can be positive, $(C, L) \OMIT{\in \gramcond}$, or negative, $\neg (C, L) \OMIT{\in \gramcond}$. 
For a configuration, \textit{cfg}, we say that $\match(\textit{cfg}, (C, L))$ holds if there exists some line $(C', L') \in \textit{cfg}$ such that $\match((C', L'), (C, L))$, while $\match(\textit{cfg}, \neg(C, L))$ holds if no such $(C', L')$ exists. For positive match conditions, parameter predicates are applied to instances of the matched context-lines, whereas negative match conditions identify omitted parameters/statements.
% The negative match conditions, on the other hand, do not contribute any matched instances and instead mainly serve to identify omitted parameters and statements (which can then be added in the repair action).

%Repair rules can also be conditioned on predicates, $\grampred$, which in addition 
\myitem{Parameter Predicates.}
%% Role of parameter predicates
Once the matching (and missing) instances of context-lines have been identified, the relationships among the parameters of the matched lines that must hold or need to be changed are expressed using parameter predicates.
%we determine the relationships among the parameters of the matched lines that must hold or need to be changed. These relationships are expressed using predicates over parameters. 
%This decision may depend on the values of parameters in the matched context-lines, which is captured by parameter predicates over the symbolic variables occurring in the matching instances. 
%
The set of predicates, $\grampred$, includes the standard equality and logical connectives, as  well as domain-specific atomic predicates over configuration parameters. These domain-specific predicates are needed for expressing fine-grained properties, possibly relating lines in the source and target configurations. 
\OMIT{
\ag{next line could be deleted}
For example, Cisco and Junos configurations differ in their interface naming conventions, so in 
% a Junos translation for the configuration given in 
\figureautorefname~\ref{fig:config_translate}, the interface "\texttt{Ethernet0/1}" is translated to "\texttt{ge-0/0/1.0}". In such cases, an operator would define a predicate $\texttt{maps\_to}(i_1{\,:\,}\text{IFACE},i_2{\,:\,}\text{IFACE})$, tailored to their specific network, to specify when $i_1$ and $i_2$ represent corresponding interfaces in Cisco and Junos. 
%As shorthand, we will use $P(L_1, L_2, \ldots)$ to denote a predicate $P$ over the parameters of context-augmented lines $L_1, L_2, \ldots$
}

%% Repair action
\myitem{Repair Actions.}
%In general, repair rules have the form $(t, a)$, where $t$ is referred to as a trigger and $a$ an action. A trigger is a logical formula over one or more configurations, conditions from $\gramcond$, and predicates from $\grampred$, while an action $a \in \gramaction$ is the corresponding repair action to take when $t$ is true. Using conditions and predicates, we define two templates for triggers that capture useful properties for configuration translation.
%In addition to identifying when a misconfiguration has occurred through match conditions and parameter predicates, rules must also describe how to fix errors. 
The fixes for errors are expressed as repair actions $a \in \gramaction$, which can be either injecting, dropping, or replacing a context-line, and
%For configuration translation, these actions are always performed on the target translation.
%
%Also, note that context-lines injected by a 
may depend on the matched instances of a match condition. 
% For instance, consider line 3 of \figureautorefname~\ref{fig:grammar_example}, which defines the prefix of the configured interface. 
\OMIT{
\ag{the example below could be deleted}
For instance, the Cisco snippet in \figureautorefname~\ref{fig:config_translate} configures the interface prefix using an IP address (\texttt{10.0.0.1}) and a mask (\texttt{255.255.255.0}). However, prefixes in Junos are represented as an IP prefix (\texttt{10.0.0.1/24}). Hence, a rule that injects the correct prefix (not shown in the figure) would need to compute a prefix, $\textit{pfx}{\,:\,}\text{PFX4}$, from a given IP address and mask extracted from the matched lines.
%, $ip{\,:\,}\text{IPV4}$ and mask, $\textit{mask}{\,:\,}\text{IPV4}$. 
This computed prefix would then be part of the context-line used in the repair action. 
%Our implementation supports such derived parameters. \sw{Is this sufficient description for derived parameters, or should there be a grammar construct for them?}
}

\myitem{Useful templates for rule application.}
% In order to maintain the correctness of rules, they must be easily interpretable and auditable, regardless of whether they are produced manually by human operators or by automated methods. If rules comprised of arbitrary match conditions, parameter predicates, and repair actions are allowed, then it could lead to rules that are overly complex. This can also make it difficult to identify incorrect, extraneous, or redundant rules, which reduces efficiency as the number of rules increases.
To ensure that rules are interpretable and auditable, we define domain-specific rule templates that target common errors observed during LLM configuration translation. For these templates, match conditions are separated into source conditions, $S$, and target conditions, $T$. 
%By varying the order in which $S$, $T$, and predicates $P$ are checked, these rule templates can express differing relational properties across source configuration and target translation.
%
%
Intuitively, rule templates identify the instances of lines in a target configuration that need repair. In the first template, called \textbf{source-conditioned rules}, we identify all lines ($SL_i$) in the source configuration that match with the source conditions $S$, and  then look for the existence of lines ($TL_j$) in the target configuration that match with $T$, and where the predicates $P$ hold over $SL_i$ and $TL_j$. \textbf{Target-conditioned rules} are defined analogously, where the roles of the source and target conditions are switched. We also define a third template for \textbf{target-only} rules, which contain only target conditions.
%
% Analogously, in the second template, called \textbf{target-conditioned rules}, we identify all lines ($TL_i$) in the target configuration that match with the target conditions $T$. Then, we look for existence of lines ($SL_j$) in the source configuration that match with source conditions $S$ and where the predicates $P$ hold over $TL_i$ and $SL_j$. 
%
%Intuitively, rule templates identify the instances of lines in a target configuration where a repair action should be applied.
%to a given target translation. 
\OMIT{
\ag{could delete the formulas to save space}
The source-conditioned, target-conditioned, and target-only templates correspond to the quantifier patterns shown below:
%\sw{explain that the formulas describe when to apply the action}
\begin{align*}
    &\forall SL_i.\; \texttt{match}(SL_i, S) \; \exists. TL_j.\; \texttt{match}(TL_j, T) \wedge P(SL_i, TL_j) \\
    &\forall TL_i.\; \texttt{match}(TL_i, T) \; \exists. SL_j.\; \texttt{match}(SL_j, S) \wedge P(SL_j, TL_i) \\
    &\forall TL_i.\; \texttt{match}(TL_i, T) \land P(TL_i)
\end{align*}
}

\myitem{Rule Templates in Action.} 
\figureautorefname~\ref{fig:config_translate} shows a common misconfiguration example caused by differing default OSPF priority values between Cisco and Junos \cite{cisco-ios-docs, junos-os-docs}. Using the source-conditioned rule template, we can define a rule to find and fix this mistake:
{\small\setlength{\jot}{1pt}
\setlength{\abovedisplayskip}{4pt}\setlength{\belowdisplayskip}{4pt}
\setlength{\abovedisplayshortskip}{4pt}\setlength{\belowdisplayshortskip}{4pt}
\begin{align*}
    S &= \{ \texttt{interface }[i_1:\text{IFACE}],\, \\
    &\,\,\,\,\,\,\,\,\,\, \neg(\texttt{interface } i_1 \,/\, \texttt{ip ospf priority } [p:\text{NUM}]) \} \\
    T &= \{ \texttt{interface }[i_2:\text{IFACE}], \\
    &\,\,\,\,\,\,\,\,\,\, \neg(\texttt{interface }i_2 \,/\, \texttt{priority } [p:\text{NUM}])\} \\
    P &= \{\texttt{maps\_to}(i_1, i_2), \textbf{true}\} \\
    a &= \actinject(\texttt{interface }i_2 \,/\, \texttt{priority }1)
\end{align*}
}
% The source-conditioned rule generated by the above targets a common issue caused by differing default OSPF priority values between the Cisco and Junos configuration languages. 
If the source defines an interface $i_1$ without an explicit priority and the target defines a matching $i_2$ (via \texttt{maps\_to} predicate) that also omits priority, the rule injects "$\texttt{priority }1$" into interface $i_2$. 

\OMIT{
\myitem{Target-Conditioned Rules: Example}. 
%Conversely to the previous source-conditioned rules, we also define a template for target-conditioned rules. Checking a target-conditioned rule requires first checking each target conditions, then evaluating the source conditions based on the matched lines in the target. 
While rules under the source-conditioned template describe how to translate the presence or absence of lines in the source configuration, they do not assert that some source conditions must be true in order for some target conditions to hold. The target-conditioned rule template fills this gap.
% Hence, while source-conditioned rules express sufficient conditions for actions, target-conditioned rules describe necessary conditions. 
% \begin{align*}
%     &\left( \bigwedge_{i = 1}^{n} \texttt{match}(\textit{tgt}, T_i) \right) \land \\
%     &\forall L_1, L_2, \ldots, L_{|S^+|} \in \textit{tgt}.  \left( \bigwedge_{i = 1}^{|T^+|} \texttt{match}(L_i, T^+_i) \right) \\
%     &\implies \bigwedge_{i = 1}^m \left( \texttt{match}(\textit{src}, S_i) \land P_i(L_1, L_2, \ldots) \right)
% \end{align*}

For instance, a common mistake is that for configurations with multiple OSPF areas and interfaces, the model will often generate an interface within the context of an area that does not match the prefix of the interface. To address this error, we define the following target-conditioned rule:
\begin{align*}
    T &= \{\texttt{area }[a:\text{AREA\_ID}] \,/\, \texttt{interface } [i_1:\text{IFACE}]\} \\
    S &= \{ \texttt{interface } [i_2:\text{IFACE}] \,/\, \\
    &\,\,\,\,\,\,\,\,\,\texttt{ip address } [ip_1:\text{IPV4}]\,[mask_1:\text{IPV4}],\\
    &\,\,\,\,\,\,\,\,\,\neg(\texttt{network } [ip_2:\text{IPV4}]\,[mask_2:\text{IPV4}] \texttt{ area } [a])\} \\
    P &= \{\texttt{maps\_to}(i_1, i_2), \\
    &\,\,\,\,\,\,\,\,\,\texttt{subnet\_match}(ip_1, mask_1, ip_2, mask_2)\} \\
    a &= \textbf{drop}(\texttt{area }[a] \,/\, \texttt{interface } [i_1])
\end{align*}
Where \texttt{subnet\_match} is a predicate that is true when the address and mask of the interface $i_2$ are matched by the network address and mask of the OSPF area $a$. This rule first checks for any interface $i_1$ included in the context of area $a$ in the target. Then, if the corresponding interface $i_2$ is defined in the source, but there is no network definition matching the subnet of $i_2$ and the area $a$, the interface $i_1$ should be dropped from the context of area $a$ in the target. Hence, this rule states that for an interface $i_1$ to be configured in an area $a$, the source configuration must define a corresponding interface whose address matches the subnet of a network in area $a$.

% \todo{add an example here if you like}\\
\myitem{Target-Only Rules: Example} Unlike the source and target-conditioned rule templates, which relate source and target configurations, target-only rules describe repairs that depend only on the target. These rules are useful for capturing vendor-specific syntax as well as standalone semantic correctness. For example, we observed that when translating configurations to Junos, the LLM would frequently define statements at incorrect locations in the configuration hierarchy. More concretely, consider the following rule:
\begin{align*}
    T &= \{\texttt{routing-options / ospf}\} \\
    P &= \{\textbf{true}\} \\
    a &= \textbf{drop}(\texttt{routing-options / ospf})
\end{align*}
This rule targets a particular common mistake, where the LLM would often incorrectly configure OSPF settings inside of a "\texttt{routing-options}" block, even though in Junos, the statement "\texttt{ospf}" can only be nested inside of a "\texttt{protocols}" block. This rule ensures that if an "\texttt{ospf}" statement is generated in an inappropriate context, namely "\texttt{routing-options}", then the incorrect line will be dropped. 
}

% \vspace{-0.5\baselineskip}
\subsection{Rule enforcement}
\label{subsec:rule_enforce}
\OMIT{
\begin{figure}[h]
\centering
\includegraphics[width=\linewidth]{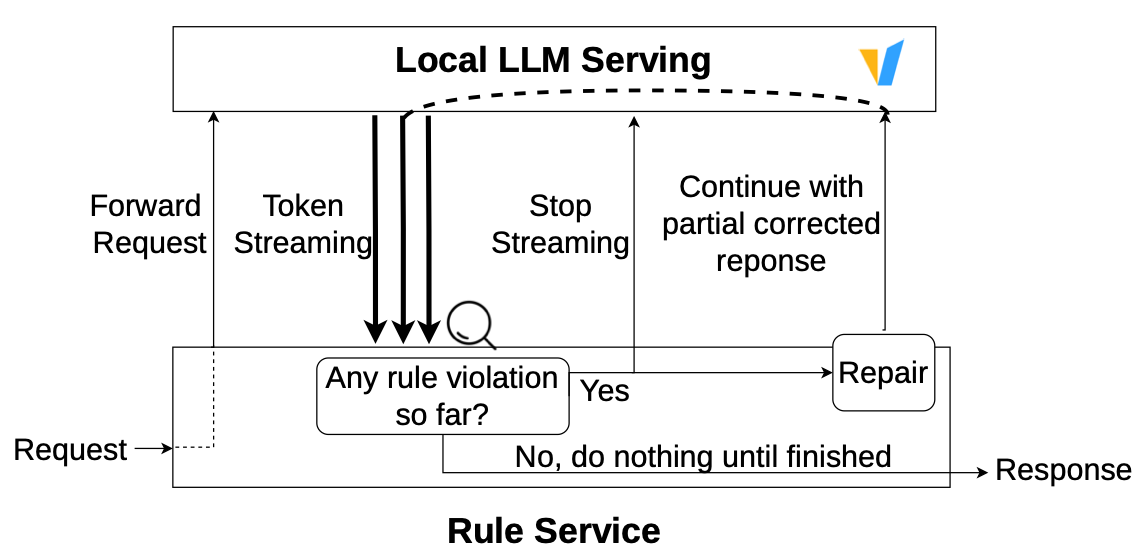}
\caption{}
\label{fig:streaming}
\end{figure}
}

We enforce rules with a streaming chunk-repair mechanism. For each network task, a lightweight Rule Service acts as a proxy between the user and the LLM serving platform: it forwards the user's request, monitors the streamed output, and returns a rule-compliant response.

% \figureautorefname~\ref{fig:streaming} shows the rule-enforcement workflow inside the Rule Service. 
The service inspects syntactically meaningful chunks, such as completed configuration blocks, as they stream and evaluates each against the rule set. On a violation, it interrupts the stream, applies the corresponding repair action, appends the repaired prefix to the original prompt, and resends the request so the LLM continues from the corrected state.

This keeps enforcement minimally invasive: the LLM and its serving stack are unchanged, and the compact rule representation guides correction without expensive token-level search. Because each retry reuses the original prompt plus the accepted prefix, serving engines with prefix/KV-cache support can reuse unchanged prefix blocks and cut regeneration cost~\cite{kvcache}. 
% Most importantly, before returning the response, the Rule Service checks and repairs the output so that it satisfies the predefined rules enforced by the system.
\section{Initial Promise}

% In this section, we highlight \sys's inital promise by using it to discover common errors made by a small LLM. We show that by guiding the small LLM to avoid these common mistakes during inference, its accuracy increases by 28X with minimal end-to-end latency degradation. 

We use \sys to discover common errors made by a small LLM, then guide the model to avoid them at inference. This raises accuracy by 28× with minimal end-to-end latency overhead.

\subsection{Setup}

% Fuzzer

We first borrow the fuzzing approach of Metha~\cite{metha} focusing on static routing and OSPF, producing independent rule-mining and test sets of  $\approx50$ and $\approx110$ configs, respectively.
%
% Emulator
To check correctness, we build a ContainerLab-based validator~\cite{containerLab}. 
For each Cisco–Junos pair, it constructs parallel star topologies with identical FRRouting~\cite{frrouting} peer routers derived from the Cisco configuration, waits for OSPF adjacencies to converge, and compares the observed routing behavior. 

\OMIT{
For each Cisco-Junos pair, we construct two parallel star-shaped topologies--a Cisco reference topology and a Junos target topology. 
Both topologies use the same FRR peer routers derived from the Cisco configuration, ensuring that the validator tests both devices under equivalent external routing environments. 
During evaluation, the validator waits for the emulated OSPF adjacencies to converge and then compares observed routing behavior. 
}
%including OSPF attributes, LSA metrics, inter-area routes, and static routes.

% To evaluate the behavioral equivalence between source configurations and candidate translations, we use emulation with a custom testbed, based on ContainerLab~\cite{}. \sw{should we mention issues with verifiers? e.g. too strict} Each testing scenario consists of separate, identical network topologies for the source and target configurations. This topology is comprised of a single router for each defined interface. For these routers, we generate minimal FRR configurations that contain only the necessary interface and area definitions to connect to the corresponding interfaces of the source or target configuration. Once router connections have been established (we consider a test failed if a connection cannot be established for a target configuration), we run a series of tests for behavioral equivalence across the source and target routers, which examine node adjacency, neighbor attributes, interface and LSA properties, and static routing. \sw{Can someone double check the networking terminology I used here?} 

% Rules

We serve  \qwenmodel~\cite{hui2024qwen2} as a small model on a single Nvidia H200 GPU instance via vLLM~\cite{vllm} and use GPT-5.5 to correct failed translations, which succeeded for all configurations in the rule-mining set.
%repeatedly propose candidate repairs when the small model's translated configuration fails validation. 
% Specifically, when a translated configuration fails validation, GPT-5.5 is prompted with the failure context and asked to propose a minimal edit to correct the misconfiguration. 
% On our training set, this repair loop achieves 100\% validation accuracy with 1.79 repair iterations on average.

% For rule generation, we first ran GPT-5.5 in a feedback loop with emulator feedback against a set of 53 Cisco configurations generates using the fuzzer and (incorrectly) translated to Junos using \qwenmodel. The GPT model was then prompted to iteratively repair each candidate translation given emulator feedback. At each iteration, the model was also instructed to only perform the minimal edit required to fix the misconfigurations reported by the emulator.
% \maria{some of the set-up is repetitve (we can say some of this in sec 2)}
We then manually mine rules from ⟨query, wrong reply, corrected reply⟩ triplets using the templates of \S~\ref{section:rule_design}, yielding 16 rules covering OSPF attributes, interface and area configuration, static routing, and Junos syntax. By enforcing these rules, $\approx85\%$ of rule-mining configurations pass validation.

\OMIT{
Finally, we mine rules by targeting the common mistakes observed during the LLM based translation and repair of the rule-mining configuration set. Given a triplet of <query, wrong reply, corrected reply>, we manually derive rules using the template described in \S~\ref{section:rule_design}.
In total, we define 16 rules covering default and explicit OSPF costs and priorities, interface configuration, OSPF area configuration, static routing, and Junos syntax. 
When enforcing these rules during translation of the rule-mining set, $\sim85\%$ of the resulting configurations pass our validation.
}
% Note that we focus on common mistakes and the mined rules turns out to achieve $\approx85\%$ accuracy when enforcing them against the rule-mining set. 
% More rules are required if targeting at 100\% accuracy.

% Rules were written manually by inspecting the errors and their corresponding fixes made by the GPT model during this loop, and generalizing common errors into the templates described in Section~\ref{section:rule_design}. This process resulted in a total of 16 rules describing translation constraints related to default and explicit values for OSPF costs and priorities, interface configuration, OSPF area configuration, static routing, and Junos syntax.

\subsection{Results}

\OMIT{
1. Mined rules 
2. Accuracy between (small model + constraint, small model only, gpt)
3. inference latency (cdf of the latency)

\minhao{More triplets don't mean having more rules. at least say how many rules order of magnitude.
More statistics on ratio of rules in templates.
How many predicates we need on average per rule.
How many matching condition ...
Table / report numbers}
}

\begin{figure}[t]
\centering
\begin{subfigure}{.90\linewidth}
    \centering
    \includegraphics[width=.99\linewidth]{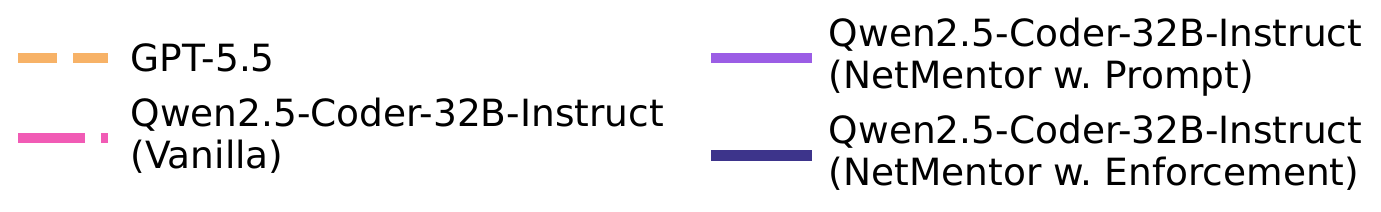}  
\end{subfigure}

\begin{subfigure}{.49\linewidth}
    \centering
    \includegraphics[width=.99\linewidth]{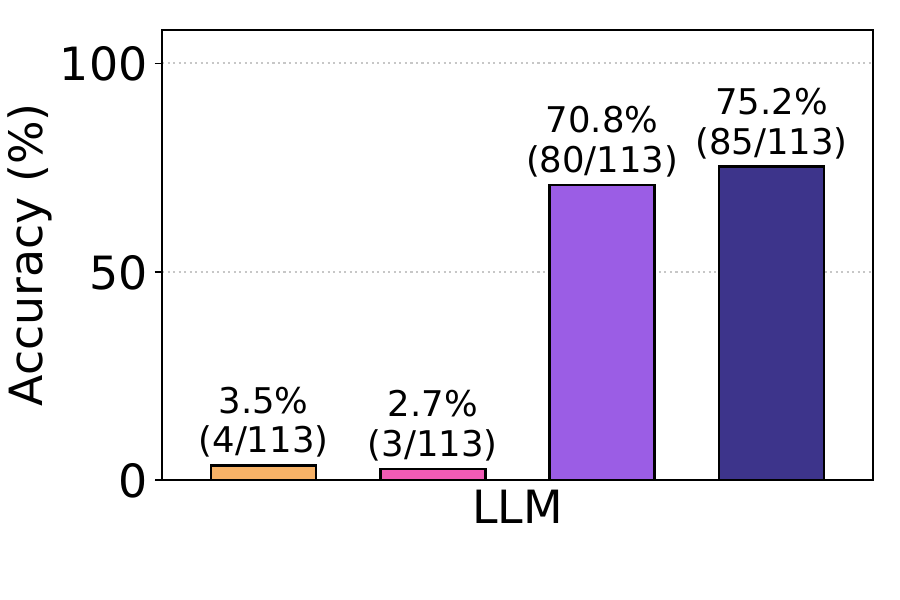}  
    \caption{Accuracy}
    \label{fig:acc}
\end{subfigure}
\begin{subfigure}{.49\linewidth}
    \centering
    \includegraphics[width=.99\linewidth]{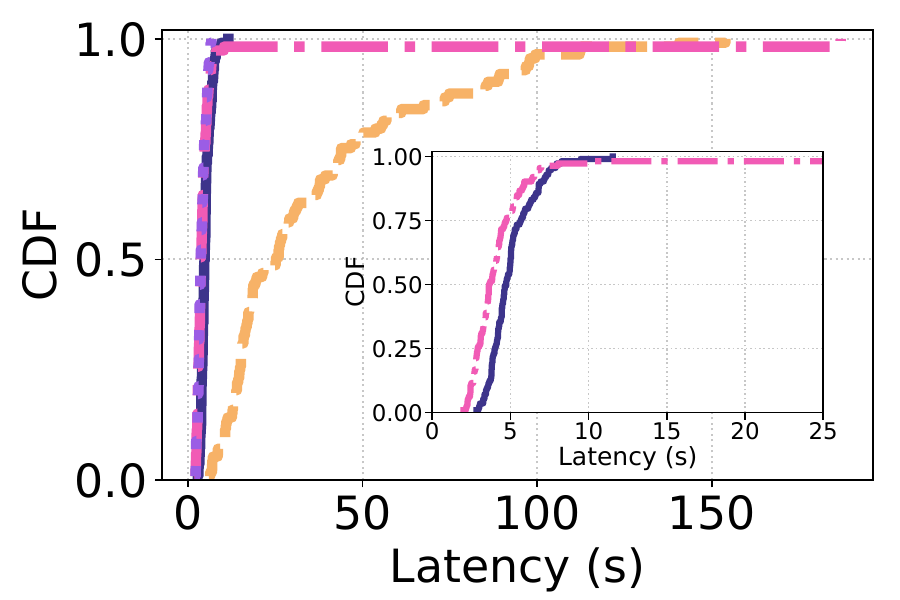}  
    \caption{Latency}
    \label{fig:latency}
\end{subfigure}
\vspace{-1em}
\caption{(a) \sys raises Qwen's accuracy by >20× over vanilla Qwen and GPT-5.5, and enforcement beats prompt injection by 6\%.
(b) \sys  cuts median latency 81\% below GPT-5.5 and shrinks the vanilla model's latency variance by two orders of magnitude.
% \ag{suggestion: change the order in the legend to: gpt, qwen-vanilla, qwen-autozoom; use this order also in (a), so auto-zoom is the right-most high column}
}
\vspace{-1em}
\end{figure}

\mypara{A few concise rules in \sys can correct many LLM errors} 
We mine 16 rules to correct many recurring translation errors: 11 source-conditioned, 3 target-conditioned, and 2 target-only, averaging 1.5 source and target conditions, 2.9 parameters, and 0.9 predicates each. We define 6 relational domain-specific predicates, relating source lines to target lines. These rules are concise and easy to audit, yet capture enough recurring structure to guide translation effectively.
\OMIT{
Our evaluations used only 16 rules in all, to correct many more translation errors during evaluations. 
Of these rules, 11 follow the source-conditioned template, 3 are target-conditioned, and 2 are target-only. The average number of source conditions (not including target-only rules) and target conditions per rule is 1.5, the average number of parameters per rule is 2.9, and the average number of predicates is 0.9. Among the predicates, we defined 6 unique domain-specific predicates, all of which are relational (i.e., they relate parameters occurring in source lines and target lines). These numbers indicate that our mined rules are concise and easy to interpret, while still providing useful corrections for effectively guiding LLM translation. 
}

\mypara{With \sys, the small LLM achieves 28× and 20× accuracy boost against the vanilla small model and frontier LLM respectively}
Both vanilla \qwenmodel and GPT-5.5 achieve low pass rates (<5\%) for the validator (\figureautorefname~\ref{fig:acc}).
\sys raises the accuracy of \qwenmodel to 75.2\% by enforcing mined rules and 70.8\% by injecting rules into the prompt.
Thus, LLM errors are not random but recurring and systematically correctable.

\OMIT{
We evaluate the accuracy for the configs generated by \qwenmodel with/without \sys as well as the GPT-5.5. \figureautorefname~\ref{fig:acc} shows that both the vanilla \qwenmodel and the frontier GPT-5.5 achieve low accuracy (<5\%) passing the validator. In contrast, \sys raises the accuracy of \qwenmodel to 75.2\% by enforcing mined rules that capture common errors and 70.8\% by simply translating rules into input prompt. 
This corresponds to at least a 20X improvement over both baselines. These results suggest that LLM errors are not entirely random; many follow recurring patterns that can be identified and systematically corrected, leading to substantial improvements in reliability.
}

\mypara{\sys's rule enforcement introduces minimal end-to-end latency overhead}
% We next evaluate the end-to-end inference latency of \qwenmodel with and without \sys, as well as that of GPT-5.5. \figureautorefname~\ref{fig:latency} shows the latency CDF over the test set.
Vanilla Qwen exhibits a pronounced long tail because for some queries, it repeatedly generates duplicated Junos configuration blocks until reaching the timeout.
\sys prevents these pathological generations, reducing latency variability by approximately two orders of magnitude while increasing the median latency by only 30\% relative to the vanilla model.
\sys achieves 81\% lower median latency while delivering substantially higher accuracy than GPT-5.5.
Overall, \sys substantially improves correctness and latency predictability while introducing only modest overhead.
\section{Future Agenda}

\myitem{Equivalence testing as a first  verification step.}
In \sys, 
%the emulator must know which properties of a candidate solution to test. 
we manually specify the properties for the emulator to test, which puts a significant burden on the user. A broader research question is how to automatically identify the task-relevant behaviors that determine whether a solution is correct, or whether two versions of behavioral specification (\eg a high-level intent and its implementation) are equivalent. This also raises a coverage question: which behaviors and corner cases have been tested, and what guarantees can we provide for solutions that pass?

\myitem{Generalizing knowledge-enforcement representations.}
 Our current rule templates generalize corrections across related examples, but the templates and rules are manually constructed through inspection and intuition. An important research direction is to automatically infer symbolic patterns, synthesize candidate rules, and evaluate their correctness and generality. One must also investigate whether rules (hence template design) should align with model tokens or remain expressed over higher-level networking objects.

\myitem{Smarter Fuzzer.}
% Rather than exploring configuration space uniformly, a smarter fuzzer should target inputs that exploit the model’s learned biases or lack of understanding. 
Rather than exploring the configuration space uniformly, a smarter fuzzer should target inputs that exploit the model's learned biases. 
% Vision models, for example, learn spurious associations from common visual patterns: because most pictures of tables include chairs, a model describing a scene might describe all scenes of tables and no chairs wrongly. Identifying this bias helps improve the model through guidance.
Vision models, for instance, learn spurious associations: because most pictures of tables include chairs, a model may wrongly infer chairs in a chairs-free scene~\cite{li-etal-2023-evaluating, singh2020dontjudgeobjectcontext, rosenfeld2018elephantroom}. Identifying such biases enables targeted guidance.
Similarly, because configurations often omit default values, a model may learn that omitted parameters can remain omitted across vendors—even when Cisco and Junos assign different OSPF priorities. Detecting such recurring assumptions could guide the fuzzer toward default-sensitive and rarely co-occurring features, exposing failures faster and reducing the cost of teaching the small model. 
\section{Conclusion}

Small open-source models are cheap but unreliable, while frontier models bring cost, latency, and fragile dependencies. We bridge this gap with a framework that discovers common mistakes, generalizes them into auditable logic rules, and uses these rules to guide a small model at inference time.
Preliminary results motivate an agenda for automating and generalizing verification, rule-mining, and data generation.
% \section*{Acknowledgements}

% This document is based on the template used for HotNets '24, modifed to use the standard ACM article class.

\bibliographystyle{ACM-Reference-Format} 
\bibliography{hotnets25-template}

\end{document}